\documentclass[12pt]{iopart}

\pdfoutput=1

\usepackage{iopams}  
\usepackage{graphicx}
\usepackage{setstack}
\usepackage{color}
\usepackage{cite}

\usepackage{braket}

\newcommand{\eqref}[1]{(\ref{#1})}
\DeclareMathAlphabet      {\mathitbf}{OML}{cmm}{b}{it}

\newcommand{\tw}{t_\mathrm{w}}

\begin{document}

\title{The out-equilibrium 2D Ising spin glass: almost, but not quite,
  a free-field theory}

\author{L.~A.~Fernandez$^{1,2}$, E.~Marinari$^{3,4,5}$, V.~Martin-Mayor$^{1,2}$, G.~Parisi$^{3,4,5}$ and  J.~J.~Ruiz-Lorenzo$^{6,2}$}

\address{$^1$ Departamento de  F\'{\i}sica Te\'orica. Facultad de Ciencias
  F\'{\i}sicas. Universidad Complutense de Madrid. Madrid 28040. Spain.}
\address{$^2$ Instituto de Biocomputaci\'on y
  F\'{\i}sica de Sistemas Complejos (BIFI), 50018 Zaragoza, Spain.}
\address{$^3$ Dipartimento di Fisica, Sapienza
  Universit\`a di Roma, I-00185 Rome, Italy.}
\address{$^4$ Nanotec, Consiglio Nazionale delle Ricerche, I-00185 Rome, Italy.}
\address{$^5$ Istituto Nazionale di Fisica Nucleare, Sezione di Roma 1, I-00185 Rome, Italy.}
\address{$^6$ Departamento de F\'{\i}sica and
  Instituto de Computaci\'on Cient\'{\i}fica Avanzada (ICCAEx), Universidad de
  Extremadura, 06071 Badajoz, Spain.}

\date{\today}

\begin{abstract}
We consider the spatial correlation function of the two-dimensional
Ising spin glass under out-equilibrium conditions. We pay special
attention to the scaling limit reached upon approaching zero
temperature.  The field-theory of a non-interacting field makes a
surprisingly good job at describing the spatial shape of the
correlation function of the out-equilibrium Edwards-Anderson Ising
model in two dimensions.
\end{abstract}

\maketitle 

\section{Introduction.}

The importance of characterizing the spatial range of spin-glass
correlations has been long recognized, both under
equilibrium~\cite{ballesteros:00,palassini:99} and out-equilibrium
conditions~\cite{fisher:88,rieger:94,marinari:96,kisker:96,joh:99,berthier:02,jonsson:02b,berthier:04,jimenez:05,berthier:05b,jaubert:07,janus:08b,aron:08,janus:09b,guchhait:14,manssen:15,janus:16,janus:17b,guchhait:17,fernandez:18a}. These
correlations may be characterized though the overlap-overlap
correlation function (for definitions, see below
Sect.~\ref{sect:observables}). However, we still lack analytical
control over the spatial shape of this correlation function, which is
a great nuisance for numerical work.

Here, we study the overlap-overlap correlation function for the Ising
spin glass in spatial dimension $D=2$ both as a function of time and
of spatial separation. Our numerical analysis is performed on lattices
large enough to be representative of the infinite system-size limit.
The two dimensional Ising spin glass undergoes a $T=0$ phase
transition, however we hope that our results would apply equally in
Ising spin glasses above its lower critical dimension (which is
believed to be $D\simeq 2.5$~\cite{boettcher:05}) in the paramagnetic phase.
In fact, recent experiments on a film
geometry~\cite{guchhait:14,guchhait:15a,guchhait:15b,guchhait:17,zhai:17}
motivated us to undertake a large scale numerical simulation of the
out-equilibrium dynamics of the $D=2$ spin
glass~\cite{fernandez:18a}. These systems, for small times, will
behave as if living in a spin glass phase, yet for larges times they will
cross over to the dynamical critical behavior of the two dimensional
Ising spin glass- a paramagnetic phase behavior. Our aim here is to present a  more
field-theoretically minded analysis of the correlation function, as
compared with our previous phenomenological
analysis~\cite{fernandez:18a}.

It came to us as a real surprise that the Langevin dynamics for the
free scalar-field makes an excellent job in describing the spatial
dependence of the spin-glass correlations. Of course, at least in an
equilibrium setting~\cite{parisi:88, amit:05}, large-distance
correlations in a paramagnetic phase (and the $D=2$ Ising spin glass
has only a paramagnetic phase for $T>0$) should be given by free-field
theory. What \emph{is} a surprise is that free-field theory is very
accurate also at short distances. Furthermore, in the large-time limit
of an equilibrated system, free-field theory can be made virtually
exact for the spin-glass through a logarithmic wave-function
renormalization (because of the vanishing anomalous
dimension~\cite{fernandez:16b}). In fact, we are able to parameterize
in a very simple way the rather heavy corrections-to-scaling found in a
previous equilibrium study~\cite{jorg:06b}.

The remaining part of this work is organized as follows. In
Sect.~\ref{sect:observables} we shall describe the model and the basic
spin-glass correlation function that we compute (for further technical
details, see Ref.~\cite{fernandez:18a}).  In Sect.~\ref{sect:C4} we
elaborate on the implications of scale invariance for the spatial
shape of the correlation function. The relationship between the
spin-glass correlations and the free-field propagator is considered in
equilibrium (Sect.~\ref{sect:eq-FF}) and out-equilibrium
(sect.~\ref{sect:non-eq-FF}).  Our conclusions are presented in
Sect.~\ref{sect:conclusions}. A number of results regarding the
free-field propagator are derived and discussed in~\ref{app:FF}.

\section{Model and Observables}\label{sect:observables}

Our dynamic variables are Ising spins, $s_\mathitbf{x}=\pm 1$, placed
in the nodes of a square lattice of linear dimension $L$. Their
interaction is given by the Edwards-Anderson
Hamiltonian~\cite{edwards:75,edwards:76} with nearest-neighbors
couplings and periodic boundary conditions
\begin{equation}\label{eq:H}
{\cal H}=-\sum_{\langle\mathitbf{x},\mathitbf{y}\rangle}
J_{\mathitbf{x},\mathitbf{y}} s_\mathitbf{x} s_\mathitbf{y} \,.
\end{equation}
We consider quenched disorder~\cite{parisi:94}, which means that the
couplings $J_{\mathitbf{x},\mathitbf{y}}$ are fixed once for all. The
couplings are drawn from the bimodal probability distribution
($J_{\mathitbf{x},\mathitbf{y}}=\pm 1$ with equal probability). Every
set $\{J_{\mathitbf{x},\mathitbf{y}}\}$ defines a \emph{sample}. We
have simulated $L=512$, which is large enough to be insensitive to the
finite size effects (see Sect.~\ref{sect:eq-FF}).  Notice that the $T=0$
phase transition is Universal (i.e. it is independent of the type of
disorder, see for example Ref.~\cite{fernandez:16b}).

Our numerical protocol is as follows. We start from a fully disordered
spin configurations (representative of infinite temperature), which is
instantaneously placed at the working temperature $T$ at the initial
time $\tw=0$.  A standard Metropolis dynamics at fixed $T$
follows. Our time unit is a full-lattice sweep, which roughly
corresponds to one picosecond~\cite{mydosh:93}. We have simulated a
multi-spin code of a $L=512$ lattice for a wide range of temperatures
($0.5\le T \le 1.1$). The number of simulated samples has been 96.
For each sample, we have run 256 replicas (for $T\ge 0.55$) or 264
replicas (for $T=0.5$).

The overlap correlation function (see Ref.~\cite{janus:09b} for a detailed discussion) is
computed from the replica-field
\begin{equation}
q^{\alpha,\beta}(\mathitbf{x},\tw) = s^{(\alpha)} (\mathitbf{x},\tw)
s^{(\beta)}(\mathitbf{x},\tw)\,,\quad (\alpha\neq\beta)\,.
\end{equation}
The $\{s^{(\alpha)} (\mathitbf{x},\tw)\}$ are \emph{real
  replicas} ($\alpha$ is the so called replica index): replicas with
different replica indices evolve under the same set of couplings
$\{J_{\mathitbf{x},\mathitbf{y}}\}$ but are otherwise statistically
independent. Hence, our correlation function is 
\begin{equation}\label{eq:C4-def}
C_4(\mathitbf{r},\tw;T) =
\overline{
\langle q^{\alpha,\beta}(\mathitbf{x},\tw) q^{\alpha,\beta}(\mathitbf{x}+\mathitbf{r},\tw)\rangle}\,,
\end{equation}
where one first take the average over the thermal noise and the
initial conditions, denoted by $\langle\ldots\rangle$. The average
over the random couplings, denoted by an overline, is only computed
afterwards. We shall restrict ourselves to displacement vectors along
one of the lattice axis [the choice between $\mathitbf{r}=(r,0)$ or
$\mathitbf{r}=(0,r)$ is immaterial, so we average over the two], and
use the shorthand $C_4(r,\tw)$~\cite{janus:09b,janus:08}.

We characterize the spatial range of correlations through the
coherence length:
\begin{equation}\label{eq:def-xi-k-kp1}
  \xi_{k,k+1}(\tw)\equiv I_{k+1}(\tw)/I_k(\tw)\,,
\end{equation}
computed by means of the integrals
\begin{equation}\label{eq:Ik-def}
  I_k(\tw)\equiv\int_0^{\infty}\mathrm{d}\,r\ r^k C_4(r,\tw)\,.
\end{equation}
Following recent
work~\cite{janus:17b,janus:09b,janus:08b,janus:16,fernandez:18a}, we
shall focus our attention in the $k=1$ length-estimate
$\xi_{12}(\tw)$.

Eventually, we have been able to equilibrate the system, in the sense
that the integrals $I_k(\tw)$ no longer depend on $\tw$ (within
errors). Of course, an infinite system never fully equilibrates.
However, in the paramagnetic phase (and spin-glasses in $D=2$ have
only a paramagnetic phase at $T>0$), we can rather think of
equilibration up to distance $r$: for any fixed distance $r$ the
$C_4(r,\tw)$ approaches its equilibrium limit $ C_4^{\mathrm{eq}}(r)$
exponentially fast in $\tw$, after a $r$-dependent time threshold is
reached, see \ref{app:assymptotics}. Given that the equilibrium
propagator decays exponentially with distance, we can regard the
system as equilibrated for all practical purposes once the
$C_4(r,\tw)$ equilibrates up to a distance (say) $r=6\,
\xi_{12}^{\mathrm{eq}}(T)$. It is therefore meaningful to study numerically
\begin{equation}
\xi_{12}^{\mathrm{eq}}(T)=\lim_{\tw\to\infty} \xi_{12}(\tw,T)\,.
\end{equation}
In our simulations, $\xi_{12}^{\mathrm{eq}}(T)$ ranges from
$\xi_{12}^{\mathrm{eq}}(T=1.1)\approx 4.3$ to
$\xi_{12}^{\mathrm{eq}}(T=0.5)\approx 39.4$: this is why we expect
that $L=512$ is large enough to accommodate $L\to\infty$
conditions~\cite{janus:08b,janus:16,fernandez:18a}.

In fact, if one takes first the limit
$L\to\infty$ and only afterwards goes to low $T$, we expect
a critical point at $T=0$:
\begin{equation}\label{eq:nu-theta-def}
\xi_{12}^{\mathrm{eq}}(T)\sim T^{-\nu}+\ldots, \quad 1/\nu=-\theta\,
\end{equation}
where the dots stand for (rather complex~\cite{fernandez:16b})
subleading corrections to scaling.  The stiffness exponent $\theta$
has been computed in a $T=0$ simulation for Gaussian-distributed
couplings, $\theta=-0.2793(3)$~\cite{khoshbakht:17} (the identity
$- \theta= 1 /\nu$, was already confirmed in former Gaussian couplings
simulations, see for example Refs.~\cite{katz:04,fernandez:16b}).  We
have checked in~\cite{fernandez:18a} that Eq.~\eqref{eq:nu-theta-def}
holds as well, with the same $\theta$, for our $J=\pm 1$ couplings.

Some readers may be unfamiliar with our coherence-length estimators,
so let us relate our $\xi_{k,k+1}$ to the second-moment correlation
length which is commonly studied in the context of equilibrium
critical phenomena~\cite{cooper:82,amit:05}. Let $\hat
C_4(\mathitbf{p},\tw)$ be the Fourier transform of
$C_4(\mathitbf{r},\tw)$.  In the thermodynamic limit $L\to\infty$, the
momentum $\mathitbf{p}$ is a continuous variable. In the presence of
rotational invariance (a reasonable assumption even for a fairly small
$\xi_{12}(\tw)$~\cite{janus:09b}), $\hat C_4$ depends on the squared
momentum $p^2$. Hence, the second moment correlation length is
\begin{equation}\label{eq:xi-second-moment-Fourier}
\xi^{\mathrm{2nd-moment}}=\sqrt{-\left.\frac{\partial \log\hat C_4}{\partial p^2}\right|_{p^2=0}}\,.
\end{equation}
Eq.~\eqref{eq:xi-second-moment-Fourier} can be conveniently adapted to
a finite lattice, hence discrete
$\mathitbf{p}$~\cite{cooper:82,caracciolo:93,amit:05}, which partly
explains its popularity. In real space, and assuming again
$L\to\infty$ and rotational invariance,
Eq.~\eqref{eq:xi-second-moment-Fourier} reads in dimension $D$
\begin{equation}\label{eq:xi-second-moment-real-space}
\xi^{\mathrm{2nd-moment},D}=\sqrt{\frac{1}{2 D}\,\frac{I_{D+1}}{I_{D-1}}}=\sqrt{\frac{\xi_{D-1,D}\,\xi_{D,D+1}}{2 D}}\,.
\end{equation}
The rationale for preferring $\xi_{12}$ over the more familiar
$\xi^{\mathrm{2nd-moment},D}$ is a practical one~\cite{janus:09b}:
statistical errors grow heavily with the index $k$ of the requested
integrals $I_k$. 

For later use, we note as well that the (equilibrium) spin-glass susceptibility 
is 
\begin{equation}\label{eq:chiSG}
\chi=\sum_{x,y=-\infty}^{\infty} C_4^{\mathrm{eq}}(x,y)\approx\ 2\pi I_1^{\mathrm{eq}}\,,
\end{equation}
where we have assumed again rotational invariance, as well as
$\xi_{12}^{\mathrm{eq}}\gg 1$, in order to approximate the double
summation by the integral $I_1^{\mathrm{eq}}$ (in general space
dimension, $\chi\propto I_{D-1}$).

\section{On the spatial structure of the correlations}\label{sect:C4}

In this section, we shall consider the Edwards-Anderson correlation function
$C_4(r,\tw;T)$ as a function of distance, temperature and time.  After
some preliminary considerations, we shall address two different
questions related with $C_4(r,\tw;T)$: (i) How the equilibrium
correlation $C_4^\mathrm{eq}(r;T)$ relates to the theory of a
free-field?  (Sect.~\ref{sect:eq-FF}); (ii) Is the out-equilibrium
correlation function $C_4(r,\tw;T)$ given by free-field theory?
(section~\ref{sect:non-eq-FF}).

Before addressing the above questions, let us frame the discussion.
An underlying assumption in our analysis is that our choice $k=1$ for
$\xi_{k,k+1}$, recall Eq.~\eqref{eq:def-xi-k-kp1}, is
immaterial~\cite{janus:08b,janus:09b}. This assumption is plausible
because scale-invariance suggests that the Edwards-Anderson correlation
function behaves for large $r$ as
\begin{equation}\label{eq:scale-invariance}
C_4(r,\tw;T)\approx\frac{1}{r^\zeta}\,
g\left[\frac{r}{l(\tw,T)},\frac{l(\tw,T)}{l_\mathrm{eq}(T)}\right]\,,\quad
l_\mathrm{eq}(T)=\lim_{\tw\to\infty} l(\tw,T)\,.
\end{equation} 
Unfortunately, we cannot extract the length scale $l(\tw,T)$ because we do not
have any \emph{a priori} information on the scaling function $g$ in
Eq.~\eqref{eq:scale-invariance}. This is why we use the integral estimators
$\xi_{k,k+1}(\tw)$, Eq.~\eqref{eq:def-xi-k-kp1}, that according to
Eq.~\eqref{eq:scale-invariance}, are proportional to $l(\tw,T)$:
\begin{equation}\label{eq:scale-invariance-xi}
\xi_{k,k+1}(\tw)=l(\tw,T)\,\frac{\int_0^\infty\mathrm{d}\,x\,x^{k+1-\zeta}g(x,\hat
  l)}{\int_0^\infty\mathrm{d}\,x\,x^{k-\zeta}g(x,\hat l)}\,,\quad \hat l=\frac{l(\tw,T)}{l_{\mathrm{eq}}(T)}\,.
\end{equation}

Eq.~\eqref{eq:scale-invariance} can be checked in the limiting case of
an equilibrated system, $\tw\to\infty$.  Indeed, because we are in a
paramagnetic phase [recall Eq.~\eqref{eq:nu-theta-def}], the
Renormalization Group predicts that the Edwards-Anderson correlations are
(asymptotically) given by the free-field
propagator~\cite{parisi:88,amit:05}
\begin{equation}\label{eq:scale-invariance-limit}
C_4^\mathrm{eq}(r;T) \sim K_0[r/\xi_\mathrm{exp}(T)]\quad\mathrm{for}\quad r\gg\xi_\mathrm{exp}(T)\,.
\end{equation}
In the above expression, which defines the so-called exponential
correlation-length $\xi_\mathrm{exp}(T)$, $K_0$ is the $0$-th order
modified Bessel function of the second kind~\cite{nist:10}. We remark
that Eq.~\eqref{eq:scale-invariance-limit} is specific for $D=2$
(see~\ref{app:FF} for general space-dimension). After making the
identification
\begin{equation}
l_\mathrm{eq}(T)\equiv \xi_\mathrm{exp}(T)\,,
\end{equation}
we see that Eq.~\eqref{eq:scale-invariance-limit} becomes a particular
case of Eq.~\eqref{eq:scale-invariance}.

\begin{figure}
\begin{center}
\includegraphics[width=\columnwidth,angle=0]{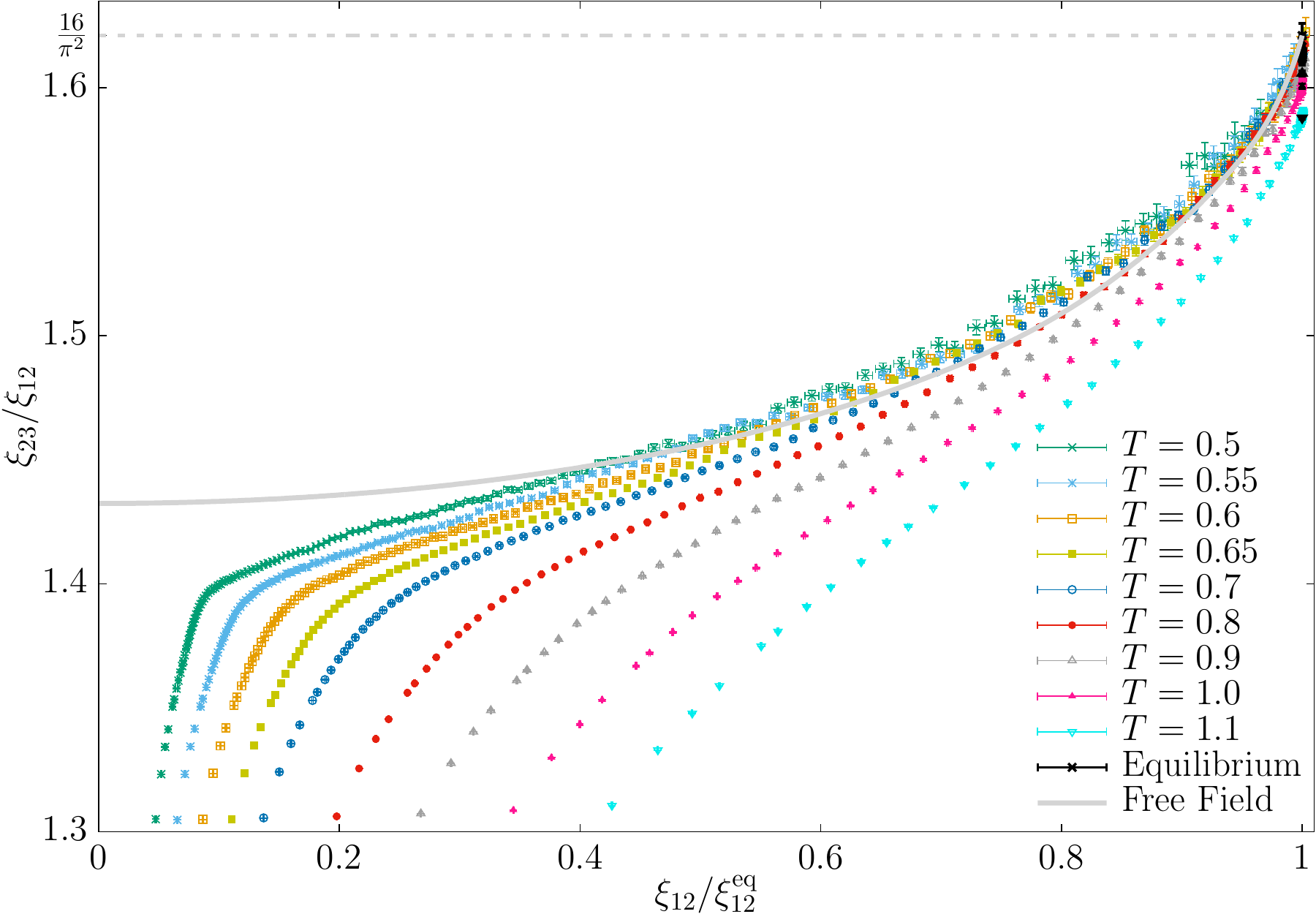}
\end{center}
\caption{As time evolves [i.e. $\xi_{12}(\tw;T)$ grows until it
    reaches its equilibrium value $\xi_{12}^{\mathrm{eq}}(T)$], the
  scale-invariant ratio $\xi_{23}(\tw,T)/\xi_{12}(\tw,T)$ varies,
  which unveils the dependency on the unknown length-scale $l(\tw)$ in
  Eqs.~(\ref{eq:scale-invariance}, \ref{eq:scale-invariance-xi},
  \ref{eq:scale-invariance-R12}). The figure shows that (barrying
  small $\xi_{12}^{\mathrm{eq}}(T)$ corrections) the temperature
  dependence can be absorbed by plotting the data as a function of the
  scale-invariant ratio
  $\xi_{12}(\tw,T)/\xi_{12}^{\mathrm{eq}}(T)$. Indeed, in agreement
  with Eq.~\eqref{eq:scale-invariance-R12}, our data collapse to a
  master curve when $\xi_{12}^{\mathrm{eq}}(T)$ grows upon lowering
  the temperature. An analogous master curve can be computed
  analytically for a non-interacting field (full line), see
  Eqs.~(\ref{eq:FF-scaling-F}) and~(\ref{eq:FF-R12})
  in~\ref{app:Integrales}. Surprisingly, the master curve for the
  free-field is a very good approximation for the Edwards-Anderson
  model. In fact, the free-field prediction might be even exact if the
  equilibrium limit $\xi_{12}(\tw,T)/\xi_{12}^{\mathrm{eq}}(T)\to 1$
  is taken first, and the scaling limit
  $\xi_{12}^{\mathrm{eq}}(T)\to\infty$ is taken afterwards.}
\label{fig:R12-2}
\end{figure}

In order to investigate further Eq.~\eqref{eq:scale-invariance},
Fig.~\ref{fig:R12-2} shows the ratio of characteristic lengths
$\xi_{23}/\xi_{12}$. Using Eq.~\eqref{eq:scale-invariance-xi} we obtain
the expected behavior of the dimensionless ratio in the scaling limit [i.e. 
$\xi_\mathrm{exp}(T)\to\infty$ at fixed $l(\tw,T)/\xi_\mathrm{exp}(T)$]:
\begin{equation}\label{eq:scale-invariance-R12}
\frac{\xi_{23}(\tw,T)}{\xi_{12}(\tw,T)}=
\frac{\left[\int_0^\infty\mathrm{d}\,x\,x^{3-\zeta}g(x,\hat
  l)\right]\left[\int_0^\infty\mathrm{d}\,x\,x^{1-\zeta}g(x,\hat
  l)\right]}{\left[\int_0^\infty\mathrm{d}\,x\,x^{2-\zeta}g(x,\hat l)\right]^2}\,,\ \hat l=\frac{l(\tw,T)}{\xi_{\mathrm{exp}}(T)}\,.
\end{equation}
The above expression unveils the role of $l(\tw,T)/\xi_\mathrm{exp}(T)$.
In fact,
should the shape of the $r$-dependence in $C_4(r,\tw;T)$ be
independent of time [thus, independent of
$l(\tw,T)/\xi_\mathrm{exp}(T)$], then also $\xi_{23}/\xi_{12}$ would
be time-independent. Instead, we see in Fig.~\ref{fig:R12-2} that
$\xi_{23}/\xi_{12}$ varies significantly as $\xi_{12}(\tw)$
grows. 

Of course, we knew beforehand that the shape of $C_4(r,\tw;T)$ \emph{must}
change with time: Eq.~\eqref{eq:scale-invariance-limit} tells us that
$C_4^\mathrm{eq}(r;T)$ decays exponentially $C_4^\mathrm{eq}(r;T)\sim
\mathrm{e}^{-r/\xi_{\mathrm{exp}}}/\sqrt{r/\xi_{\mathrm{exp}}}$. Instead,
the general arguments in~\ref{app:assymptotics} imply a
super-exponential decay for the out-equilibrium correlation function,
$C_4(r,\tw;T)\sim\mathrm{e}^{-(r/{\hat\xi})^\beta}\,,$
with $\beta>1$. What Fig.~\ref{fig:R12-2} tells us is that the change in
the functional form of $C_4(r,\tw;T)$ happens gradually.

However,  \emph{there is} something surprising in the large-$\tw$ limit
in Fig.~\ref{fig:R12-2}. Barring high-temperature corrections, the
equilibrium $\xi_{23}/\xi_{12}$ turns out to be compatible with
$16/\pi^2$, which is its free-field value \eqref{eq:FF-R12}. This is
the first indication suggesting that
Eq.~\eqref{eq:scale-invariance-limit} might work for
$r\leq\xi_{\mathrm{exp}}$ as well, way before its natural validity
range.

Let us now find a workaround on the annoying dependence on
$l(\tw,T)/\xi_\mathrm{exp}(T)$ in Eq.~\eqref{eq:scale-invariance-xi}
(this dependency is a nuisance because, although $\xi_\mathrm{exp}(T)$
can be obtained from our data, see Sect.~\ref{sect:eq-FF}, $l(\tw)$
remains a mystery). Fortunately, Eq.~\eqref{eq:scale-invariance-xi}
suggests that the (computable) dimensionless ratio
$\xi_{12}(\tw,T)/\xi_{12}^{\mathrm{eq}}(T)$ is a one to one function
of $l(\tw,T)/\xi_\mathrm{exp}(T)$. Hence, we can compare
out-equilibrium data at different temperatures by plotting
$\xi_{23}/\xi_{12}$ as a function of
$\xi_{12}(\tw,T)/\xi_{12}^{\mathrm{eq}}(T)$, see
Fig.~\ref{fig:R12-2}. Barring corrections for small
$\xi_{12}^{\mathrm{eq}}(T)$ it is clear that the data collapse to a
master curve, which is exactly what we expect from
Eq.~\eqref{eq:scale-invariance-xi}. We note as well that the same
curve can be computed analytically for the free-field (full curve in
Fig.~\ref{fig:R12-2}). The free-field master curve turns out to be
fairly close to the limiting master curve for the Edwards-Anderson
model.

We are now ready to address the questions posed at
the beginning of this Section.

\section{The equilibrium Edwards-Anderson correlations and the theory of a free-field}\label{sect:eq-FF}

Let us consider the paramagnetic phase of a typical $D$-dimensional
spin system in thermal equilibrium. The asymptotic behaviors of the
correlation function are
\begin{equation}\label{eq:scaling-standard}
C^\mathrm{eq}(r\ll\xi_\mathrm{exp})\sim\frac{1}{r^{D-2+\eta}}\,,\quad C^\mathrm{eq}(r\gg\xi_\mathrm{exp})\sim\xi_\mathrm{exp}^{D-2-\eta}\,\frac{K_Q(r/\xi_\mathrm{exp})}{(r/\xi_\mathrm{exp})^Q}\,,
\end{equation}
where $\eta$ is the anomalous dimension, $Q=(D-2)/2$ and $K_Q$ is the
$Q$-th order modified Bessel function of the second
kind~\cite{nist:10}. The normalizations in
Eq.~\eqref{eq:scaling-standard} ensure that (i)
$C^\mathrm{eq}(r=1)\sim 1$ [which is certainly the case for the
  Edwards-Anderson $C_4^{\mathrm{eq}}(r;T)$], and (ii) the asymptotic
behavior for small and large $r$ connect smoothly at
$r=\xi_{\mathrm{exp}}$.\footnote{The $r\gg\xi_\mathrm{exp}$ asymptotic
  behavior in Eq.~\eqref{eq:scaling-standard} has an additional factor
  $\xi^{-\eta}$ as compared with the free-field,
  Eq.~\eqref{eq:FF-G-eq-limit}. This extra factor is the origin of the
  wave-function renormalization
  $Z_\phi\sim\xi^{\eta/2}$~\cite{parisi:88,amit:05}, which for $\eta=0$ will produce a logarithmic divergence,
  see also the discussion
  of Eq. (\ref{eq:scaling-non-standard}).}

However, let us take seriously for one minute the suggestion  that the
large-distance asymptotic behavior holds all the way down to $r\sim 1$.
 Now,  specializing to $D=2$ and recalling that
$K_0(y\to0)\sim\mathrm{log}\, 1/y\,$, we see that the condition $C^\mathrm{eq}(r=1)\sim 1$ implies that
\begin{equation}\label{eq:scaling-non-standard}
C^\mathrm{eq}_{2D,\mathrm{non-standard}}(r)\sim \frac{K_0(r/\xi_\mathrm{exp})}{\mathrm{log}\,\xi_\mathrm{exp}}\,.
\end{equation}
Funnily enough, Fig.~\ref{fig:R12-2} suggests that
the (equilibrium) 2D Ising spin-glass could really follow the
non-standard behavior in Eq.~\eqref{eq:scaling-non-standard}, even for $r<\xi_{\mathrm{exp}}$ . Our aim
here will be exploring further this hypothesis.

Eq.~\eqref{eq:scaling-non-standard} suggests to start by fitting our
equilibrium correlation function to
\begin{equation}\label{eq:non-standard-fit}
C_4^\mathrm{eq}(r;T)={\cal A}(\xi_\mathrm{exp}) \left[\,K_0\Big(\frac{r}{\xi_{\mathrm{exp}}(T)}\Big)\ +\ K_0\Big(\frac{L-r}{\xi_{\mathrm{exp}}(T)}\Big)\,\right]\,,\quad L=512,
\end{equation}
where ${\cal A}(\xi_\mathrm{exp})$ is an amplitude depending on temperature
through $\xi_\mathrm{exp}(T)$. We have included in~\eqref{eq:non-standard-fit}
the first-image term, $K_0[(L-r)/\xi_\mathrm{exp}]$ (mind our periodic
boundary conditions), as a further control of finite-size effects. In fact,
results turn out to vary by less than a tenth of an error bar (one standard deviation)  when the image
term is removed. This agreement confirms that the $L=\infty$ limit has been
effectively reached.

\begin{table}[t]
\centering
\begin{tabular*}{\columnwidth}{@{\extracolsep{\fill}}crrrcll}
\br
$T$ & $r_\mathrm{min}$ & $r_\mathrm{max}$ & $\chi^2/\mathrm{dof}$ & $A(\xi_\mathrm{exp})$ & $\xi_\mathrm{exp}$ & $\xi_{12}^{\mathrm{eq}}/\xi_{\mathrm{exp}}$\\
\mr
0.50 & 19 &  202 & 13.72/182 & 0.2295 (34) & 24.98 (30)     &  1.5758 (27) \\ 
0.55 & 14 &  166 & 70.53/151 & 0.2469 (27) & 18.10 (16)   &  1.5757 (24) \\ 
0.60 & 10 &  142 & 36.85/131 & 0.2655 (20) & 13.63 (8)    &  1.5771 (21) \\ 
0.65 &  8 &  113 & 47.37/104 & 0.2812 (19) & 10.68 (6)    &  1.5772 (16) \\ 
0.70 &  8 &  103 & 62.60/94  & 0.2981 (20) & \ 8.59 (4)   &  1.5815 (22) \\ 
0.80 &  5 &   66 & 22.82/60  & 0.3259 (12) & \ 5.942 (17) &  1.5798 (9) \\ 
0.90 &  3 &   50 & 35.97/46  & 0.3566 (7) & \ 4.358 (8)  &  1.5841 (5) \\ 
1.00 &  4 &   39 & 5.38/34  & 0.3867 (10) & \ 3.355 (6)  &  1.5893 (8) \\ 
1.10 &  4 &   31 & 9.70/26  & 0.4189 (11) & \ 2.671 (4)  &  1.5994 (9) \\
\br
\end{tabular*}

\caption{\label{tab:fit_K0} For each temperature in our simulations,
  we report the results of a fit to Eq.~\eqref{eq:non-standard-fit}.
  Given that the numerical estimates of $C_4^{\mathrm{eq}}(r;T)$ are
  dramatically correlated for different distances $r$, we use as fit's
  figure of merit, the diagonal $\chi^2$ (i.e. the $\chi^2$ statistics
  as computed keeping only the diagonal terms in the covariance
  matrix). These correlations are responsible for the anomalously low
  $\chi^2$ that we find. The distances included in the fit are
  $r_\mathrm{min}\leq r\leq r_\mathrm{max}$ (see
  Ref.~\cite{fernandez:18a} for details).  To compute errors in the
  fit parameters, namely ${\cal A}(\xi_\mathrm{exp})$ and
  $\xi_\mathrm{exp}$, we employ the jackknife as implemented
  in~\cite{yllanes:11}: we fit for each jack-knife block (using for
  all blocks the diagonal covariance matrix), and compute errors from
  the blocks fluctuations. We also report the ratio
  $\xi_{12}^{\mathrm{eq}}/\xi_{\mathrm{exp}}$ (in order to account for
  statistical correlations, errors were computed with the
  jackknife). In a free-field theory,
  $\xi_{12}^{\mathrm{FF},\mathrm{eq}}/\xi_{\mathrm{exp}}=\pi/2=1.5707963\ldots$,
  see Eq.~\eqref{eq:FF-xi-k-kp1-eq}, which is fairly close to our
  numerical results for the Edwards-Anderson model. The behavior of
  $\xi_{12}^{\mathrm{eq}}/\xi_{\mathrm{exp}}$ in the limit of large
  $\xi_{\mathrm{exp}}$ is studied in
  Fig.~\ref{fig:scaling-eq}--bottom.  }
\end{table}

\begin{figure}
\begin{center}
\includegraphics[width=\columnwidth,angle=0]{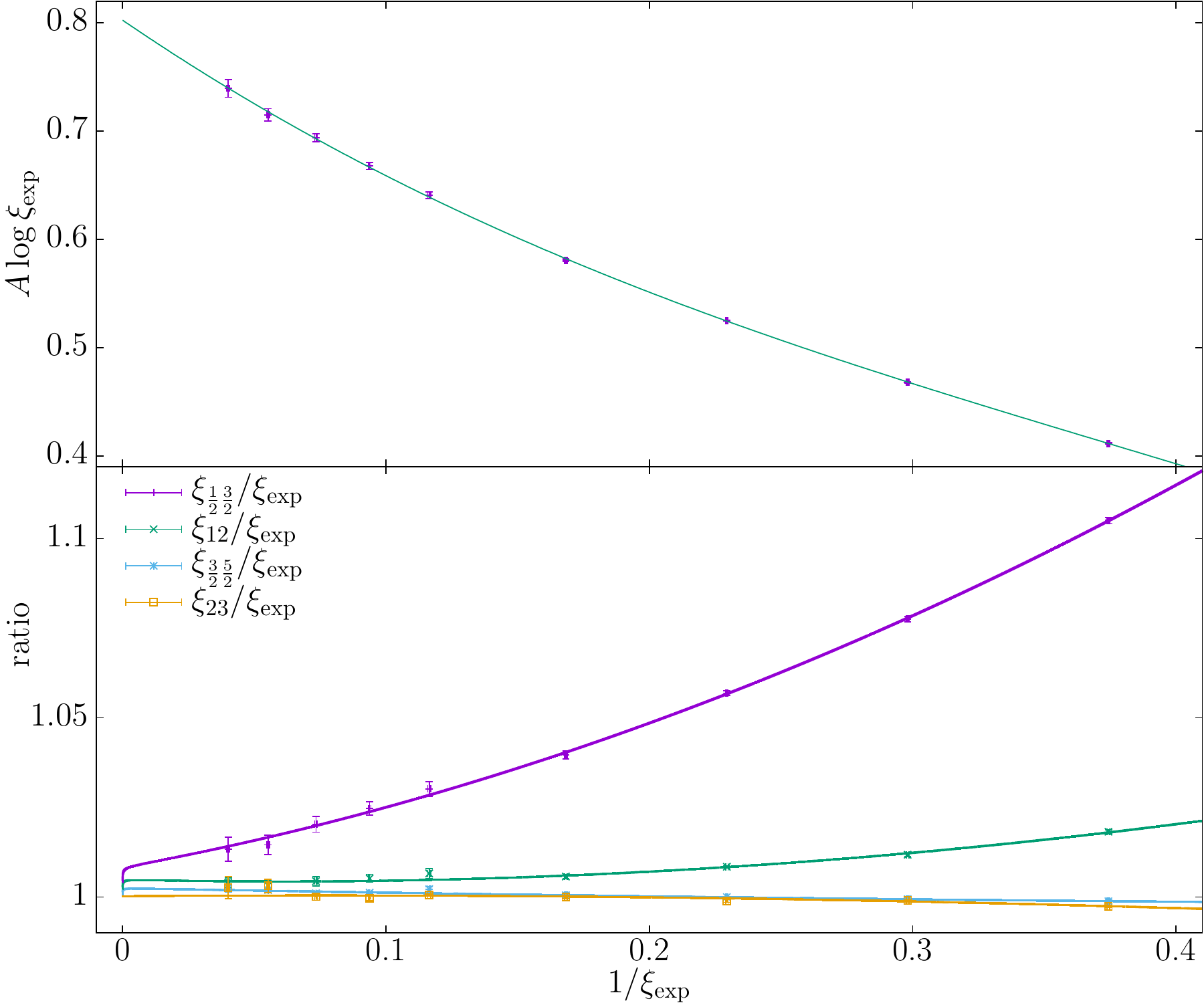}
\end{center}
\caption{{\bf Top}: The consistency condition $C^\mathrm{eq}(r=1)\sim
  1$, see Eq.~\eqref{eq:scaling-non-standard}, requires the
  amplitude in the fit in Eq.~\eqref{eq:non-standard-fit} to scale as
  ${\cal A}(\xi_\mathrm{exp}\to\infty)\sim
  1/\mathrm{log}\,\xi_\mathrm{exp}$.  Indeed, the plot shows that
  ${\cal A}(\xi_\mathrm{exp}) \mathrm{log}\, \xi_\mathrm{exp}$ is
  excellently represented, $\chi^2/\mathrm{dof}=2.2/5$, by a cubic
  polynomial in $1/\xi_\mathrm{exp}$, implying ${\cal
    A}(\xi_\mathrm{exp}) \mathrm{log}\, \xi_\mathrm{exp}\approx 0.8$
  for large $\xi_\mathrm{exp}$. {\bf Bottom:} if the non-standard
  scaling~\eqref{eq:scaling-non-standard} holds true, all the $J_k$
  defined in Eq.~\eqref{eq:Jk-def} should tend to 1 when
  $\xi_\mathrm{exp}\to\infty$ (the $J_k$ are the Edwards-Anderson
  $\xi^{\mathrm{eq}}_{k,k+1}/\xi_{\mathrm{exp}}$ divided by their
  free-field counterparts). We show $J_k$ as a function of
  $1/\xi_\mathrm{exp}$, for $k=1/2$, $1$,$3/2$ and $2$. Lines are fits
  to Eq.~\eqref{eq:short-distances-4} [note that the function
    $v(\xi_\mathrm{exp})$, Eq.~\eqref{eq:short-distances-3}, is
    continuous, but has infinite slope at
    $\frac{1}{\xi_\mathrm{exp}}=0$]. The corresponding figures of
  merit of these are $\chi^2/\mathrm{dof}=2.4/6$ ($k=1/2$),
  $\chi^2/\mathrm{dof}=4.0/6$ ($k=1$), $\chi^2/\mathrm{dof}=3.9/6$
  ($k=3/2$) and $\chi^2/\mathrm{dof}=10.0/6$ ($k=2$).}
\label{fig:scaling-eq}
\end{figure}

The results of the fit to Eq.~\eqref{eq:non-standard-fit} are reported
in Table~\ref{tab:fit_K0}. As the reader may check, even in the most
difficult case, namely $T=0.5$, $\xi_\mathrm{exp}(T)$ is computed with
1\% accuracy.  We find as well, see Fig.~\ref{fig:scaling-eq}--top,
that the consistency condition $C^\mathrm{eq}(r=1)\sim 1$ expressed in
Eq.~\eqref{eq:scaling-non-standard} is well satisfied by our data.

A further confirmation of Eq.~\eqref{eq:non-standard-fit} comes from the second-moment correlation length. Combining Eq.~(\ref{eq:xi-second-moment-real-space}), as applied to $D=2$,  with Eq.~\eqref{eq:FF-xi-k-kp1-eq}) we see that
Eq.~\eqref{eq:non-standard-fit} implies
\begin{equation}\label{eq:caramba}
\xi^{\mathrm{2nd-moment},\mathrm{eq}}=\xi_{\mathrm{exp}}\,.
\end{equation}
Thanks to previous results in Ref.~\cite{fernandez:16b}, we may compare
these two characteristic lengths, see Tables~\ref{tab:fit_K0} and~\ref{tab:second_moment}. The agreement is most satisfactory.

\begin{table}[b]
\centering
\begin{tabular*}{0.3\columnwidth}{@{\extracolsep{\fill}}cl}
\br
$T$ & $\xi^{\mathrm{2nd-moment},\mathrm{eq}}$\\
\mr
0.50  & 23.99(17)\\
0.55  & 17.95(11)\\
0.65  & 10.753(39) \\
0.60  & 13.712(65)\\
0.70  & \ 8.649(26)\\
0.80  & \ 5.968(13)  \\
0.90  & \ 4.3854(71) \\
1.00  & \ 3.3657(45) \\
1.10  & \ 2.6782(51)\\
\br
\end{tabular*}
\caption{\label{tab:second_moment} Second moment correlation-lenght in
  equilibrium as computed in an $L=128$ system by means of a Parallel
  Tempering simulation (data from Ref.~\cite{fernandez:16b}). We
  expect $\xi^{\mathrm{2nd-moment},\mathrm{eq}}=\xi_{\mathrm{exp}}$,
  see the discussion of Eq.~\eqref{eq:caramba}.  In fact, letting
  aside $T=0.5$ (because a $L=128$ lattice is clearly too small to
  represent the $L\to\infty$ limit for that temperature), the
  agreeement with the corresponding values for $\xi_\mathrm{exp}$ in
  Table~\ref{tab:fit_K0} is impressive.}
\end{table}

Of course, one cannot expect Eq.~\eqref{eq:non-standard-fit} to hold for all
$r$. Indeed, the fit works only for $r\geq r_\mathrm{\min}$, see
Table~\ref{tab:fit_K0}. We find that the ratio
$r_\mathrm{\min}/\xi_{\mathrm{exp}}$ is small, but remains finite as
$\xi_{\mathrm{exp}}$ grows upon lowering $T$. In fact, we have empirically
found that
\begin{equation}\label{eq:short-distances-1}
C_4^{\mathrm{eq}}(r;T)-{\cal A}(\xi_{\mathrm{exp}})K_0(r/\xi_{\mathrm{exp}})=
{\cal B}(\xi_{\mathrm{exp}})\frac{\mathrm{exp}[-7 r/\xi_{\mathrm{exp}}]}{(r/\xi_{\mathrm{exp}})^{0.2}}\,.
\end{equation}
We have checked at $T=0.5$ and $0.55$ that
Eq.~\eqref{eq:short-distances-1}, for which we lack a theoretical
justification, works for all $r\geq 1$ (in the sense of an acceptable
$\chi^2/\mathrm{dof}$). Our standard regularity condition $C^{\mathrm{eq}}_4(r=1,T)\sim 1$ tells us that
\begin{equation}\label{eq:short-distances-2}
{\cal B}(\xi_{\mathrm{exp}})\sim \left[\frac{1}{\xi_{\mathrm{exp}}}\right]^{0.2}\quad \mathrm{for}
\quad \xi_{\mathrm{exp}}\to\infty\,.
\end{equation}

We are finally ready to consider the extrapolation to large
$\xi_{\mathrm{exp}}$ of the ratios
$\xi^{\mathrm{eq}}_{k,k+1}/\xi_{\mathrm{exp}}$. We shall start by
dividing the $\xi^{\mathrm{eq}}_{k,k+1}/\xi_{\mathrm{exp}}$ by their
free-field value in Eq.~\eqref{eq:FF-xi-k-kp1-eq}:
\begin{equation}\label{eq:Jk-def}
J_k\equiv\frac{\xi^{\mathrm{eq}}_{k,k+1}}{\xi_{\mathrm{exp}}}\,
\frac{1}{2}\,\frac{\Gamma^2[(k+1)/2]}{\Gamma^2[(k+2)/2]}\,.
\end{equation}
Our working hypothesis is that $J_k\to 1$ for large
$\xi_{\mathrm{exp}}$· Then, a straightforward computation starting
from Eqs.~(\ref{eq:short-distances-1},\ref{eq:short-distances-2})
predicts that the finite-$\xi_{\mathrm{exp}}$
corrections for
$\xi_{k,k+1}^{\mathrm{eq}}=I_{k+1}^{\mathrm{eq}}/I_k^{\mathrm{eq}}$
take the form of a series-expansion in the corrections-to-scaling function
\begin{equation}\label{eq:short-distances-3}
v(\xi_\mathrm{exp})=\frac{1}{(\xi_{\mathrm{exp}})^{0.2}\,\cal{A}(\xi_{\mathrm{exp}})}\,.
\end{equation}
Besides, we have the standard corrections in $1/\xi_{\mathrm{exp}}$,
stemming from our considering continuous functions of
$r/\xi_{\mathrm{exp}}$ while numerical data can be obtained only for
integer $r$. Accordingly, we have fitted our data to
\begin{equation}\label{eq:short-distances-4}
J_k=1\ +\  a_k\, v(\xi_\mathrm{exp})\ +\ \frac{b_k^{(1)}}{\xi_\mathrm{exp}}\ +\ 
\frac{b_k^{(2)}}{\xi^2_\mathrm{exp}}\,,
\end{equation}
with fitting parameters $a_k$, $b_k^{(1)}$ and $b_k^{(2)}$.  We have
found fair fits to Eq.~\eqref{eq:short-distances-4}, see
Fig.~\ref{fig:scaling-eq}--bottom, even for $k$ as small as $k=1/2$
(the smaller the $k$, the more highlighted the small-$r$ region).
${\cal A}(\xi_{\mathrm{exp}}$) is promoted to a continuous function of
$\xi_{\mathrm{exp}}$ through the fit in Fig.~\ref{fig:scaling-eq}--top
[this is needed to compute $v(\xi_{\mathrm{exp}})$].  To assess the
relative importance of the correction terms in
Eq.~\eqref{eq:short-distances-4}, we may consider $b_k^{(1)}/a_k$: in
Fig.~\ref{fig:scaling-eq}-bottom, these ratios of amplitudes are
$b_{1/2}^{(1)}/a_{1/2}\approx 40$, $b_{1}^{(1)}/a_{1}\approx 6.6$,
$b_{3/2}^{(1)}/a_{3/2}\approx 11$, and $b_{2}^{(1)}/a_{2}\approx 52$.

\begin{figure}
\begin{center}
\includegraphics[width=\columnwidth,angle=0]{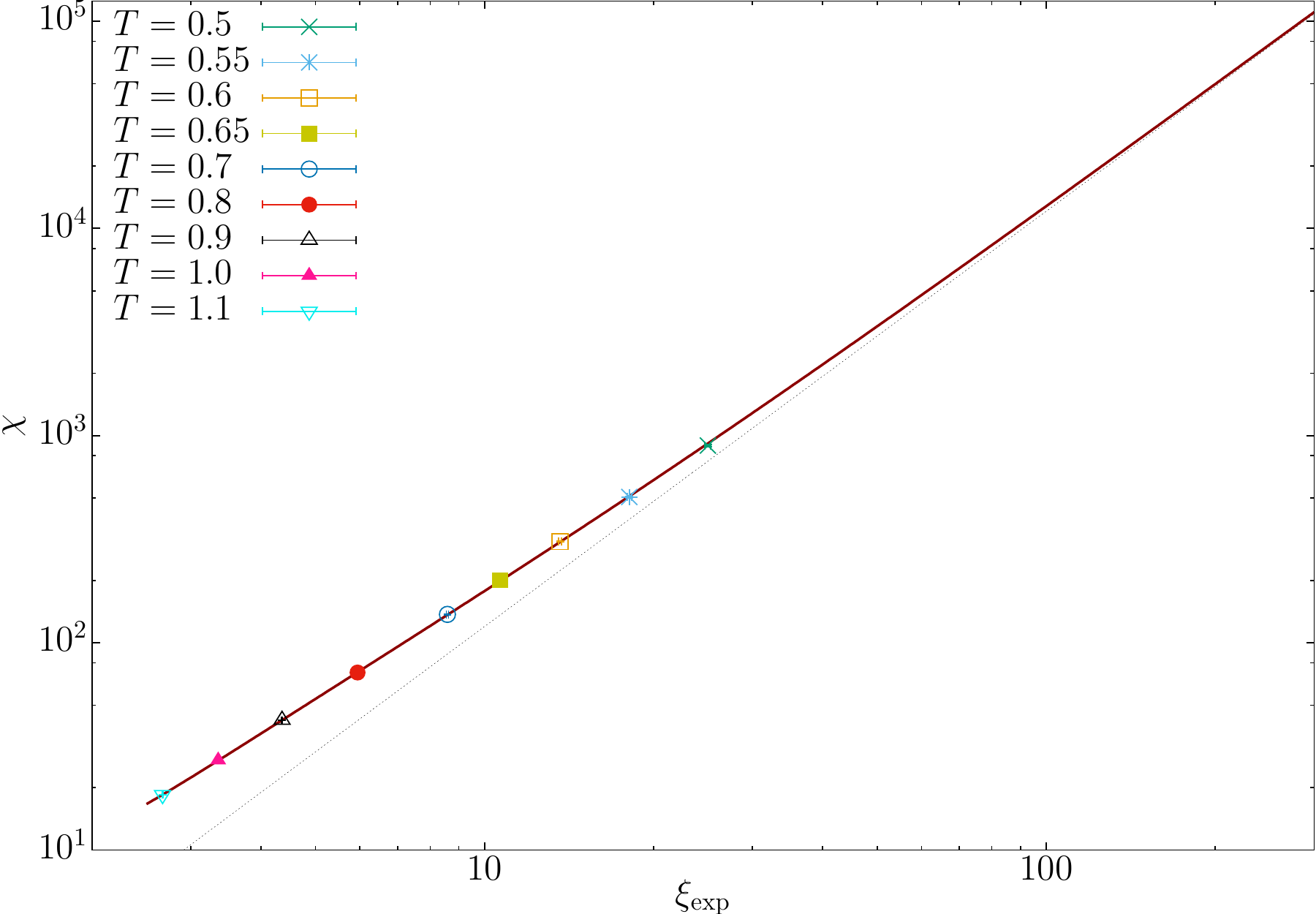}
\end{center}
\caption{Equilibrium spin glass susceptibility, $\chi=2\pi
  I_1^{\mathrm{eq}}$, see Eq.~\eqref{eq:chiSG}, as a function of
  $\xi_\mathrm{exp}$. The full line is a fit to
  Eq.~\eqref{eq:short-distances-5} ($\chi^2/\mathrm{dof}=2.8/5$). One
  could set $b_\chi^{(3)}=0$ in Eq.~\eqref{eq:short-distances-5} but
  at the prize of including only data with $\xi_\mathrm{exp}>4$ in the
  fit (in such a case, one finds $\chi^2/\mathrm{dof}=3.6/4$). The
  dotted line is the dominant term in
  Eq.~\eqref{eq:short-distances-5}, $\chi\sim
  b_\chi^{(0)}\xi_\mathrm{exp}^2$. The horizontal and vertical ranges
  of the plot have been chosen to match those of Fig. 2 in
  Ref.~\cite{jorg:06b}.}
\label{fig:suscept}
\end{figure}

 Notice that the equilibrium second-moment correlation length was
 computed in Ref.~\cite{jorg:06b} [which coincides with
   $\xi_\mathrm{exp}$, see Eq.~\eqref{eq:caramba} and
   Table~\ref{tab:second_moment}], as well as the spin-glass
 susceptibility, recall Eq~\eqref{eq:chiSG}. A very large value
 $\xi_\mathrm{exp}\approx 200$ was reached thanks to a combination of
 Parallel Tempering, cluster methods and Finite-Size
 Scaling~\cite{jorg:06b}. However, the scaling of $\chi$ was barely
 under control, in spite of the very large $\xi_\mathrm{exp}$. The
 short-distances behavior identified in
 Eqs.~(\ref{eq:short-distances-1},\ref{eq:short-distances-2}) explains
 this difficulty. Indeed, using the equivalence $\chi=2\pi
 I_1^{\mathrm{eq}}$, only valid in $D=2$, one easily finds that
\begin{equation}\label{eq:short-distances-5}
\chi=\xi^2 \Big[b_\chi^{0}\ + a_\chi\, v(\xi_\mathrm{exp})\ +\ \frac{b_\chi^{(1)}}{\xi_\mathrm{exp}}\ +\ 
\frac{b_\chi^{(2)}}{\xi^2_\mathrm{exp}}\ +\ 
\frac{b_\chi^{(3)}}{\xi^3_\mathrm{exp}}\ +\ \ldots\ \Big]\,,
\end{equation}
where $a_\chi$ and the $b_\chi^{(i)}$ are scaling amplitudes. A fair
fit to Eq.~\eqref{eq:short-distances-5} is shown in the full 
line in Fig.~\ref{fig:suscept}. The width of that full line has been
chosen to correspond with the error bars, while the dotted line in
Fig.~\ref{fig:suscept} is the leading term $\chi\sim
b_\chi^{(0)}\xi_\mathrm{exp}^2$. We see in Fig.~\ref{fig:suscept} that
the full and the dotted lines coalesce only for
$\xi_\mathrm{exp}>100$, in nice agreement with the results found in
Ref.~\cite{jorg:06b}.

In summary, in the scaling limit $\xi_{\mathrm{eq}}\to\infty$, the
equilibrium correlation-function for the Ising spin glass seems to
follow the non-standard scaling in
Eq.~\eqref{eq:scaling-non-standard}. However, some readers may
consider far-fetched our parameterization of short-distances
corrections to the free-field propagator in
Eqs.~(\ref{eq:short-distances-1}) and
~(\ref{eq:short-distances-2}). These skeptical readers may keep the
more conservative conclusion that violations to the free-field
prediction $J_k(\xi_{\mathrm{eq}}\to\infty)=1$ are, at most, of 0.3\%
for $k=1/2, 1, 3/2$, and $2$.

\section{The out-equilibrium Edwards-Anderson correlations and the theory of a free-field}\label{sect:non-eq-FF}

Relating the Langevin dynamics of a free-field with the spin-glass dynamics
may seem surprising at first sight. Indeed, the dynamics of a spin-glass in its
paramagnetic phase may be characterized through a scaling function~\cite{fernandez:18a}
\begin{equation}\label{eq:scaling-F}
\frac{\xi_{12}(\tw,T)}{\xi_{12}^\mathrm{eq}(T)} ={\cal F}\left(\frac{t_w}{\tau(T)}\right)\ +\ {\cal O}\Big(\, [\xi_{12}(\tw,T)]^{-\omega},[\xi_{12}^{\mathrm{eq}}(T)]^{-\omega}\,\Big)\,,
\end{equation}
where exponent $\omega$ controls corrections to scaling, $\tau(T)$ is a characteristic time scale, and  the dynamics
at short times is described by a dynamic exponent $\hat z$:
\begin{equation}\label{eq:hat-z-def}
{\cal F}(x\to 0)\propto x^{1/\hat z}\,\,.
\end{equation}
We have found empirically $\hat z\approx 7$
for the Edwards-Anderson model~\cite{fernandez:18a}. 

\begin{figure}
\begin{center}
\includegraphics[width=\columnwidth,angle=0]{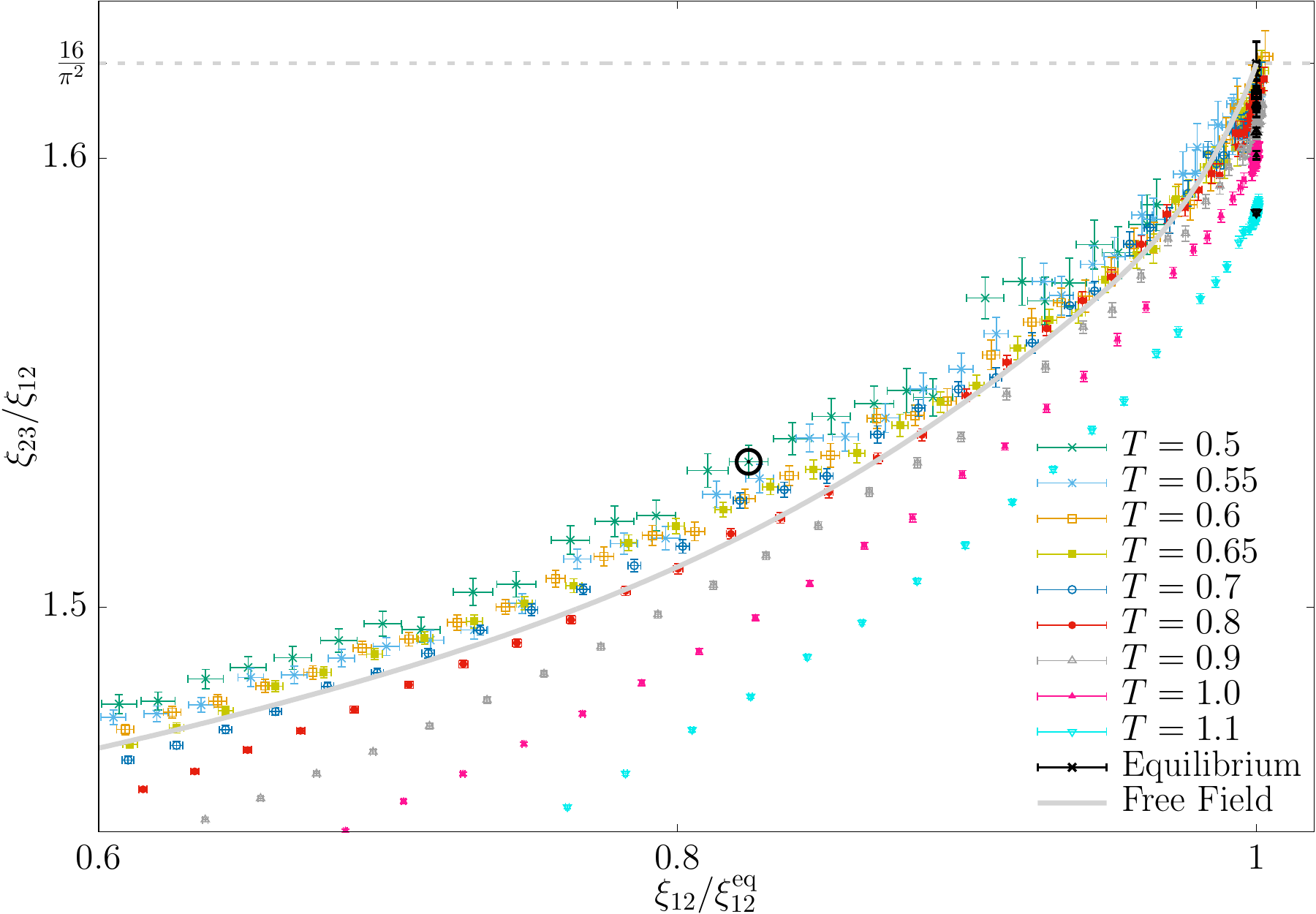}
\end{center}
\caption{Zoom of data in Fig.~\ref{fig:R12-2}. The scaling limit
  $\xi_{\mathrm{exp}}(T)\to\infty$ at fixed
  $\xi_{12}(\tw,T)/\xi_{12}^{\mathrm{eq}}(T)$ slightly differs for
  the Edwards-Anderson model (data points) and for the non-interacting
  field (full line). However, disentangling the two-models behavior
  becomes difficult upon approaching equilibrium,
  $\xi_{12}(\tw,T)/\xi_{12}^{\mathrm{eq}}(T)\to 1$. The time $\tw$
  which is explicitly compared with the free-field model in
  Fig.~\ref{fig:compara-C4} is marked by a circle (for $T=0.5$).}
\label{fig:R12-3}
\end{figure}

The analogous exponent for the free-field is $\hat z_{\mathrm{FF}}=2$
(\ref{app:Integrales}). The obvious, hardly surprising conclusion is
that spin-glass dynamics is enormously slower than free-field
dynamics. However, one may~\emph{synchronize clocks} between these two
wildly differing systems by requiring (superscripts FF stand for
\emph{free field})
\begin{equation}\label{eq:two-clocks} 
f=\frac{\xi_{12}(\tw,T)}{\xi_{12}^\mathrm{eq}(T)}=
\frac{\xi^{\mathrm{FF}}_{12}(\tw^{\mathrm{FF}})}{\xi_{12}^{\mathrm{FF},\mathrm{eq}}}\,.
\end{equation}
This clock synchronization was implicitly performed in Fig.~\ref{fig:R12-2}. 
We zoom this figure in Fig.~\ref{fig:R12-3} making it clear that the clock-synchronization
works only approximately: the free-field and the Edwards-Anderson limit behaves
in the same way only in the limit of a system in thermal equilibrium.

\begin{figure}
\begin{center}
\includegraphics[width=\columnwidth,angle=0]{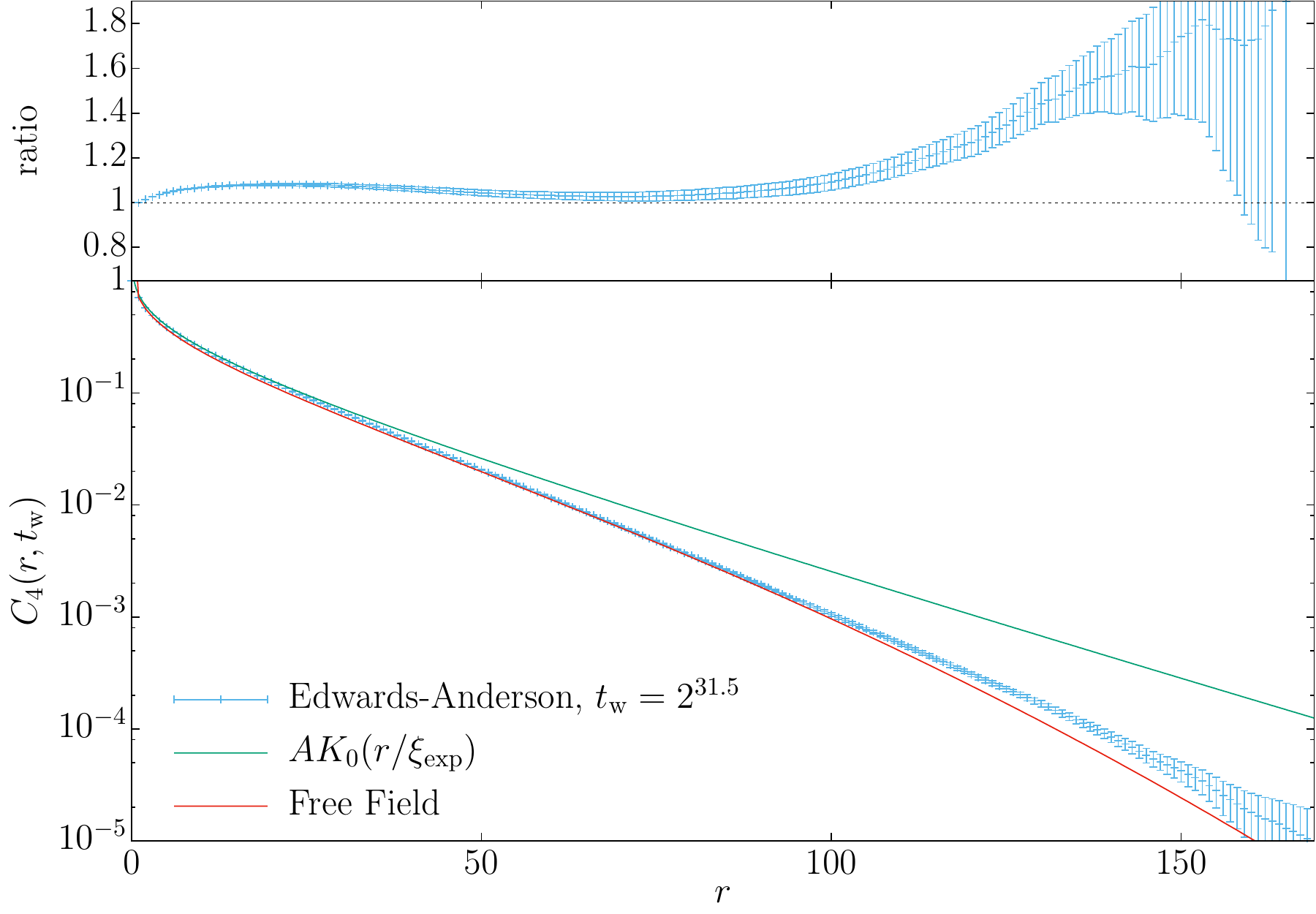}
\end{center}
\caption{{\bf Main:} For $T=0.5$, we compare the Edwards-Anderson
  correlation function $C_4(r,\tw)$ with its free-field counterpart
  $G(r,\tw^{\mathrm{FF}})$, see
  Eq.~\eqref{eq:FF-representacion-integral}. To match the parameters
  for the free-field, we fix
  $\xi^{\mathrm{FF}}_{12}(\tw^{\mathrm{FF}})/\xi_{12}^{\mathrm{FF},\mathrm{eq}}=0.824585$,
  the value pinpointed by the circle in Fig.~\ref{fig:R12-3}, and take
  $\xi_\mathrm{exp}(T=0.5)$ from Table~\ref{tab:fit_K0}. The overall
  normalization of the free-field is chosen to have
  $C_4(r=1,\tw)=G(r=1,\tw^{\mathrm{FF}})$. We also compare the two
  propagators with the asymptotic form of the equilibrium
  Edwards-Anderson correlation function, ${\cal A}(\xi_\mathrm{exp})
  K_0(r/\xi_\mathrm{exp})$.  {\bf Top}: ratio
  $C_4(r,\tw)/G(r,\tw^{\mathrm{FF}})$ as a function of $r$.  }
\label{fig:compara-C4}
\end{figure}

In order to further expose the difference, in
Fig.~\ref{fig:compara-C4} we compare the Edwards-Anderson model
correlation function $C_4(r,\tw;T)$ with its free-field counterpart in
Fig.~\ref{fig:compara-C4}, after the appropriate parameter matching.
It is clear that, even setting the same $\xi_\mathrm{exp}$ for both
models and synchronizing the clocks as in Eq.~\eqref{eq:two-clocks},
the free-field propagator has a higher curvature, as a function of $r$.

\section{Conclusions}\label{sect:conclusions}

We have studied the out-of-equilibrium dynamics of the two dimensional
Edwards-Anderson model with binary couplings. We have been able to
study the full range of the dynamics: from the initial transients to
the equilibrium through numerical simulations with a time span of 11 orders
of magnitude. We have considered the spatial dependence of the
Edwards-Anderson correlation function $C_4(r,\tw)$, that has been
compared with the propagator of a free-field theory. Much to our
surprise, we found that, after an appropriate \emph{clock
  synchronization} between the two models, the free-field propagator
provides a very good approximation to $C_4(r,\tw)$ in the
out-equilibrium regime. Furthermore, in the scaling limit
$\xi_{12}^{\mathrm{eq}}$ for the equilibrium regime, after a
logarithmic wavefunction renormalization, we find extremely difficult
to distinguish the two models numerically.

\section{Acknowledgments}

We thank Prof. Gabriel \'Alvarez for discussions.  This project has
received funding from the European Research Council (ERC) under the
European Union’s Horizon 2020 research and innovation program (grant
agreement No 694925). We were partially supported by MINECO (Spain)
through Grant Nos. FIS2015-65078-C2, FIS2016-76359-P, by the Junta de
Extremadura (Spain) through Grant No. GRU10158, GR18079 and IB16013
(these five contracts were partially funded by FEDER). Our simulations
were carried out at the BIFI supercomputing center (using the
\emph{Memento} and \emph{Cierzo} clusters) and at ICCAEx supercomputer
center in Badajoz (\emph{Grinfishpc} and \emph{Iccaexhpc}). We thank
the staff at BIFI and ICCAEx supercomputing centers for their
assistance.

\appendix

\section{The out-equilibrium dynamics of the free scalar field}\label{app:FF}

The Edwards-Anderson model in spatial dimension $D=2$ lies within its
paramagnetic phase at all positive temperatures. Therefore, the relevant
Renormalization-Group fixed point is the one of the free scalar-field (see
e.g.~\cite{parisi:88,amit:05}). This observation implies that, at least in
equilibrium, the free-field fixed point rules the system behavior at distances
$r\gg \xi_\mathrm{eq}$.

However, the $D=2$ Edwards-Anderson model and the free-field theory
\emph{might} differ for distances $r\sim \xi_\mathrm{eq}$. Futhermore, at these
length-scales, the two theories should be compared both under equilibrium and
out-equilibrium conditions. In order to confront the two models, we compute
here for the free-field the same quantities that were studied for the
Edwards-Anderson model in the main text.

Our starting point is the Langeving dynamics for a
free field~\cite{parisi:88}. At the initial time, the field is fully
disordered. The two-body correlation function $G(\mathitbf{r},\tw)$
is the analogous in the free-field theory of the Edwards-Anderson correlation
function $C_4(\mathitbf{r},\tw)$, recall Eq.~\eqref{eq:C4-def}. We can
compute explictly the free-field $G$ in Fourier space
\begin{equation}\label{eq:FF-Fourier-Space}
  \hat G(\mathitbf{p},\tw)=\frac{\big(1-\exp[-2\tw (p^2+\xi_{\mathrm{exp}}^{-2})]\big)}{p^2+\xi_{\mathrm{exp}}^{-2}}\,.
\end{equation}
The above expression defines the so-called \emph{exponential} correlation
length, $\xi_{\mathrm{exp}}$ (indeed, $\hat G(\mathitbf{p},\tw)$ tends to the
Gaussian propagator $1/(p^2+\xi_{\mathrm{exp}}^{-2})$ in the limit of large
$\tw$). Note as well that there are two characteristic lengths in
Eq.~\eqref{eq:FF-Fourier-Space}, namely the correlation length
$\xi_{\mathrm{exp}}$ and the diffusion length $\sqrt{\tw}$. Thus, before
starting our computation, it will be useful to introduce dimensionless
length ($\mathitbf{y}$) and time variables ($w$):
\begin{equation}\label{eq:y-w-def}
\mathitbf{y}=\mathitbf{r}/\xi_{\mathrm{exp}}\,,\quad w=2\tw/\xi_{\mathrm{exp}}^2\,.
\end{equation}
Rotational-invariance implies that the propagator will depend only on the
length $y$ of vector $\mathitbf{y}$ (on a lattice, rotational invariance is
recovered only in the continuum limit
$\xi_{\mathrm{exp}}\to\infty$~\cite{parisi:88}; in the context of out-equilibrium
spin glasses, the recovery of rotational invariance was investigated
in~\cite{janus:09b}).

A straightforward computation (\ref{app:calculillo}) allows us to transform back
Eq.~\eqref{eq:FF-Fourier-Space} from Fourier to real space:
\begin{equation}\label{eq:FF-representacion-integral}
G(\mathitbf{r},\tw)=\xi_{\mathrm{exp}}^{2-D}\,F_D(y,w)\,,\ 
F_D(y,w)=\frac{1}{(4\pi)^{D/2}}\int_0^w\mathrm{d}s\,\frac{\mathrm{exp}[-s -\frac{y^2}{4s}]}{s^{D/2}}\,.
\end{equation}
Armed with Eq.~\eqref{eq:FF-representacion-integral} we can compute (the
free-field analogous of) the $I_k(\tw)$ integrals defined in
Eq.~\eqref{eq:Ik-def}. This computation is performed
in~\ref{app:Integrales}. Eq.~\eqref{eq:FF-representacion-integral} makes it
simple as well the discussion of the large $y$ limit taken at fixed $w$
(\ref{app:assymptotics}). 

The opposite limit, $w\to\infty$ for fixed $y$,
yields the (equilibrium) Gaussian propagator 
(see~\cite{parisi:88} for further details):
\begin{equation}\label{eq:FF-G-eq-limit}
G^{\mathrm{eq}}(\mathitbf{r})=\frac{\xi_{\mathrm{exp}}^{2-D}}{(4\pi)^{D/2}}\int_0^\infty\,\mathrm{d}s\,\frac{\mathrm{exp}[-s -\frac{y^2}{4s}]}{s^{D/2}}=\frac{\xi_{\mathrm{exp}}^{2-D}}{(2\pi)^{D/2}} \frac{K_Q(y)}{y^Q}\,,
\end{equation}
where $Q=(D-2)/2$ and $K_Q$ is the $Q$-th order modified Bessel function of
the second kind~\cite{nist:10}. The large and small-$y$ behavior for
$D>2$ are
\begin{equation}\label{eq:FF-eq-large-y}
G^{\mathrm{eq}}(\mathitbf{r}/\xi_{\mathrm{exp}}\to\infty)\sim
\frac{\mathrm{e}^{-y}}{y^{(D-1)/2}}\,,\quad 
G^{\mathrm{eq}}(\mathitbf{r}/\xi_{\mathrm{exp}}\to 0)\sim \frac{1}{y^{D-2}}\,.
\end{equation}
The neighborhood of $y\to 0$ for the case $D=2$ deserves special care:
\begin{equation}
G^{\mathrm{eq}}(\mathitbf{r}/\xi_{\mathrm{exp}}\to 0)\sim \mathrm{log} \frac{1}{y}\,.
\end{equation}

\subsection{Integral estimators of dynamic correlations}\label{app:Integrales}

In analogy with Eq.~\eqref{eq:Ik-def}, we shall characterize the free-field
propagator through its moments (the superindex $\mathrm{FF}$ stands for \emph{free
  field})
\begin{equation}
I^{\mathrm{FF}}_k(\tw) =\int_0^\infty\mathrm{d}r\, r^k\, G(r,\tw)\,,
\end{equation}
where we have exploited the isotropy of the free-field propagator. We shall
specialize to $D=2$, and compute the moments for a propagator of the form  
\begin{equation}
G(r,\tw)={\cal A} \int_0^w\,\mathrm{d}s\,\frac{\mathrm{exp}[-s -\frac{y^2}{4s}]}{s}\,,
\end{equation}
recall Eqs.~(\ref{eq:y-w-def},\ref{eq:FF-representacion-integral}). In
particular, Eq.~\eqref{eq:FF-representacion-integral} 
implies for the amplitude ${\cal A}=1/(4\pi)$. However, the main results in
this section will be ${\cal A}$-independent (in particular, ${\cal A}$ could
depend on $\xi_\mathrm{exp}$ or $w$). We find
\begin{eqnarray}
I^{\mathrm{FF}}_k(\tw)&=& {\cal A}\, \xi_{\mathrm{exp}}^{k+1}\,\int_0^{\infty}\mathrm{d}y\,
y^k\int_0^w\,\mathrm{d}s\,\frac{\mathrm{exp}[-s -\frac{y^2}{4s}]}{s}\,,\\ 
&=&  {\cal A}\, \xi_{\mathrm{exp}}^{k+1}\, 2^k\,\Gamma[(k+1)/2]\, \gamma[(k+1)/2,w]\,,
\end{eqnarray}
by interchanging the ordering of the $y$ and $s$ integrals. In the above
expression, $\Gamma(x)$ is Euler's Gamma function and $\gamma(x,w)$ is the
lower incomplete Gamma function
\begin{equation}\label{eq:gamma-def}
\gamma(x,w)=\int_0^w\mathrm{d}s\,
s^{x-1}\mathrm{e}^{-s}\,,\quad
\gamma(x,w\to \infty)=\Gamma(x)+{\cal O}(w^{x-1} \mathrm{e}^{-w})\,.
\end{equation}
For later use, we recall its small-$w$ behavior:
\begin{equation}\label{eq:gamma-w-assymptotics}
\gamma(x,w\to 0)=\frac{w^x}{x}+{\cal O}(w^{x+1})\,,\quad 
\end{equation}

The $\xi^{\mathrm{FF}}_{k,k+1}(\tw)$ estimate of the size of the coherence length, recall
Eq.~\eqref{eq:def-xi-k-kp1}, is
\begin{equation}\label{eq:FF-xi-k-kp1}
\xi^{\mathrm{FF}}_{k,k+1}(\tw)\equiv\frac{I^{\mathrm{FF}}_{k+1}(\tw)}{I^{\mathrm{FF}}_k(\tw)}=
2\,\frac{\Gamma[(k+2)/2]}{\Gamma[(k+1)/2]}\,
\frac{\gamma[(k+2)/2,w]}{\gamma[(k+1)/2,w]}\,\xi_\mathrm{exp}\,.
\end{equation}
The equilibrium limit, $w\to\infty$, is approached exponentially in $w$
[Eq.~\eqref{eq:gamma-def}]:
\begin{equation}\label{eq:FF-xi-k-kp1-eq}
\xi^{\mathrm{FF},\mathrm{eq}}_{k,k+1}= 2\,\frac{\Gamma^2[(k+2)/2]}{\Gamma^2[(k+1)/2]}\, \xi_\mathrm{exp}\,.
\end{equation}
In other words, the integral estimators of the coherence-length, in
equilibrium but also out-equilibrium (at fixed $w$), are proportional to the
exponential correlation length $\xi_\mathrm{exp}$.

In the main text, we payed a major attention to the approach to equilibrium
of $\xi_{12}$ as computed in the Edwards-Anderson model. The free-field
analogous of Eq.~\eqref{eq:scaling-F} is
\begin{equation}\label{eq:FF-scaling-F}
\frac{\xi^{\mathrm{FF}}_{12}(\tw)}{\xi_{12}^{\mathrm{FF},\mathrm{eq}}}=\frac{\gamma(3/2,w)}{\Gamma(3/2)}
\frac{\Gamma(1)}{\gamma(1,w)}\,.
\end{equation}
It is remarkable that Eq.~\eqref{eq:FF-scaling-F} conforms \emph{exactly} to
the ansatz expressed for the Edwards-Anderson model in
Eq.~\eqref{eq:scaling-F}. Furthermore, because
$w=2\tw/\xi^2_{\mathrm{exp}}(T)$, we find
$\tau_{\mathrm{FF}}(T)=\xi^2_\mathrm{exp}(T)/2$ for the free-field analogous of the
time scale in Eq.~\eqref{eq:scaling-F}.

We can also compute the free-field exponent $\hat z_{\mathrm{FF}}$, recall
Eq.~\eqref{eq:hat-z-def}, from the small-$w$ expansion of
Eq.~\eqref{eq:FF-scaling-F} [recall Eq.~\eqref{eq:gamma-w-assymptotics}]:
\begin{equation}
\frac{\xi^{\mathrm{FF}}_{12}(\tw)}{\xi_{12}^{\mathrm{FF},\mathrm{eq}}}=\frac{2\sqrt{\pi}}{3}w^{1/2}+{\cal O}(w^{3/2})\,,
\end{equation}
which implies for the free-field $\hat z_{\mathrm{FF}}=2$.  The reader
may check from
Eqs.~(\ref{eq:gamma-w-assymptotics},\ref{eq:FF-xi-k-kp1},\ref{eq:FF-xi-k-kp1-eq})
that the small-$w$ behavior is
$\xi^{\mathrm{FF}}_{k,k+1}(\tw)/\xi^{\mathrm{FF},\mathrm{eq}}_{k,k+1}\sim
\sqrt{w}$ for any $k$, hence the result $\hat z_{\mathrm{FF}}=2$ is
$k$-independent. Because $\hat z_{\mathrm{FF}}=2$ is rather smaller
than the $\hat z\approx 7$ value that we found numerically for the
Edwards-Anderson model, we conclude that the dynamics for the
Edwards-Anderson is enormously slower than the free-field Langevin
dynamics, which is hardly surprising.

Nevertheless, Eq.~\eqref{eq:FF-scaling-F} shows that
$\xi^{\mathrm{FF}}_{12}(\tw)/\xi_{12}^{\mathrm{FF},\mathrm{eq}}$ is a
monotonously increasing function of $w$. Hence, one can parameterize
the free-field dynamics in terms of
$\xi^{\mathrm{FF}}_{12}(\tw)/\xi_{12}^{\mathrm{FF},\mathrm{eq}}$,
rather than $w$. In this way, we can obtain a meaningful comparison of
the free-field with the Edwards-Anderson dynamics. The quantities
compared are dimensionless ratios such as
$\xi_{k,k+1}(\tw)/\xi_\mathrm{exp}$ [its value for the free-field is
  given in Eq.~\eqref{eq:FF-xi-k-kp1}], or in terms of ratios not
involving $\xi_{\mathrm{exp}}$ such as $\xi_{23}(\tw)/\xi_{12}(\tw)$,
recall Figs.~\ref{fig:R12-2}.  From Eq.~\eqref{eq:FF-xi-k-kp1}, we
easily find
\begin{equation}\label{eq:FF-R12}
\frac{\xi^{\mathrm{FF}}_{23}(\tw)}{\xi^{\mathrm{FF}}_{12}(\tw)}=\frac{4}{\pi}\frac{\gamma(2,w)\gamma(1,w)}{\gamma^2(3/2,w)}\,.
\end{equation}
The limiting values are $\xi^{\mathrm{FF}}_{23}/\xi^{\mathrm{FF}}_{12}=9/(2\pi)$ (for $w\to 0$), and
$\xi^{\mathrm{FF}}_{23}/\xi^{\mathrm{FF}}_{12}=16/\pi^2$ (for $w\to\infty$).

\subsection{Asymptotic behavior of $F_D(y,w)$ (large $y$ at fixed $w$)}\label{app:assymptotics}

For any finite fixed-time $\tw$, the free-field propagator in Fourier space,
$\hat G(\mathitbf{p},\tw)$ see Eq.~\eqref{eq:FF-Fourier-Space}, is an analytic
function in the whole complex-plane of the variable $p^2$. It follows that the
function $F_D(y,w)$, defined in Eq.~\eqref{eq:FF-representacion-integral},
tends to zero at large $y$ faster than $\mathrm{e}^{-A y}$ for any $A>0$ (a
simply exponential decay corresponds with a pole singularity at
$p^2=-A^2$~\cite{parisi:88}). This statement is in apparent contradiction with
the asymptotic behavior in Eq.~\eqref{eq:FF-eq-large-y} which is exact, but
only for $\tw=\infty$. The way out of the paradox is simple: $\hat
G(p^2=-\xi_\mathrm{exp}^2,\tw)=2\tw$ which becomes a pole singularity only in
the $\tw\to\infty$ limit. It is clear that, at finite $\tw$, some sort of
crossover phenomenon is present. In this section we aim to discuss this
crossover.

We start from the integral representation~\eqref{eq:FF-representacion-integral}
\begin{eqnarray}
F_D(y,w)&=&\frac{1}{(4\pi)^{D/2}}\int_0^w\,\mathrm{d}s\,\mathrm{e}^{\Psi_D(s,y)}\,,\label{eq:IRPsi}\\
\Psi_D(s,y)&=& -\frac{y^2}{4s}\, -s\, -\frac{D}{2}\,\mathrm{log}\, s\,.
\end{eqnarray}
Consider the function $\Psi_D(s,y)$ at fixed $y$. $\Psi_D(s,y)$ tends to
$-\infty$ both for $s\to 0,\infty$. These two asymptotic behaviors of
$\Psi_D(s,y)$ are separated by a maximum at
\begin{equation}\label{eq:s-star}
s^*(y)=\frac{y^2}{D+\sqrt{D^2+4y^2}}\,.
\end{equation}
Note that $s^*(y\to 0)\sim y^2$, but $s^*(y\to \infty)\sim y/2$. 

Now, imagine that we hold $y$ fixed ($y$ should be large enough to have
$s^*(y)\approx y/2$ to a good approximation). If $w\gg s^*(y)$ we can estimate
$F_D(y,w)$ through a straightforward saddle-point expansion around $s^*(y)$
that reproduces the $w=\infty$ asymptotic behavior in
Eq.~\eqref{eq:FF-eq-large-y}:
\begin{equation}\label{eq:expansion-1}
F^{(\mathrm{SP})}_D(y,w)\sim \frac{\mathrm{e}^{-y}}{y^{(D-1)/2}}\,.
\end{equation}
The error induced by the finite $w$ is $\sim
\mathrm{e}^{\Psi_D(w,y)}/|\partial_w \Psi_D(w,y)|$, hence exponentially small.

However, because $s^*(y)\approx y/2$ for large $y$, upon increasing $y$ the
saddle point $s^*(y)$ eventually exits the integration interval $0<s<w$
(i.e. for $y\gtrsim 2w$ we have $s^*(y)>w$). Obviously, the saddle-point
expansion becomes inaccurate for such a large $y$. Under such circumstances,
the integrand in Eq.~\eqref{eq:IRPsi} is maximal at $s=w$, which gives the
large-$y$ expansion
\begin{equation}\label{eq:expansion-2}
F^{(\mathrm{Extreme})}_D(y,w)\sim\frac{\mathrm{e}^{\Psi_D(y,w)}}{|\partial_w
  \Psi_D(w,y)|}=\frac{\mathrm{exp}[-w-\frac{y^2}{4w}]}{w^{D/2}}\frac{4
  w^2}{y^2-2Dw- 4 w^2}\,.
\end{equation}

In summary, for any (dimensionless) time variable $w$ one may identify a
(dimensionless) crossover length $l_\mathrm{co}$ through
$s^*(l_\mathrm{co})=w$. If $y\ll l_\mathrm{co}$ then $F_D(y,w)$ is given to an
excellent accuracy by its equilibrium limit, Eq.~\eqref{eq:FF-G-eq-limit}.
Instead, for $y\gg l_\mathrm{co}$ the asymptotic behavior is given by
Eq.~\eqref{eq:expansion-2}. Eq.~\eqref{eq:s-star} provides asymptotic
estimates for the cross-over length,
\begin{equation}
l_\mathrm{co}(w\to\infty) \sim 2w\,,\quad\mathrm{and}\quad
l_\mathrm{co}(w\to 0) \sim \sqrt{2Dw}\,.
\end{equation}

\subsection{Back to real space: the computation of $F_D(y,z)$}\label{app:calculillo}
For the sake of completeness, let us sketch the derivation of
Eq.~\eqref{eq:FF-representacion-integral}. We need to perform the inverse
Fourier-transform:
\begin{equation}
G(\mathitbf{r},\tw)=\int \frac{\mathrm{d}^D \mathitbf{p}}{(2\pi)^D}\,\mathrm{e}^{\mathrm{i}\mathitbf{p}\cdot\mathitbf{r}}\,\frac{\big(1-\exp[-2\tw (p^2+\xi_{\mathrm{exp}}^{-2})]\big)}{p^2+\xi_{\mathrm{exp}}^{-2}}\,.
\end{equation}
 After introducing the
(dimensionless) length and time variables
$y$ and $w$, recall Eq.~\eqref{eq:y-w-def}, as well as  the dimensionless
 momentum $\mathitbf{u}\equiv \mathitbf{p}\,\xi_{\mathrm{exp}}$, we find
\begin{equation}
G(\mathitbf{r},\tw)=\xi_{\mathrm{exp}}^{2-D} F_D(y,w)\,,\ 
F_D(y,w)=\int \frac{\mathrm{d}^D \mathitbf{u}}{(2\pi)^D}\,\mathrm{e}^{\mathrm{i}\mathitbf{u}\cdot\mathitbf{y}}\frac{1-\mathrm{e}^{-w(u^2+1)}}{u^2+1}\,.
\end{equation}
Next, we note that the derivative with respect to $w$ of $F_D(y,w)$ can be
computed by derivating under the integral sign (we are left with a Gaussian
integral):
\begin{equation}\label{eq:derivada-integral}
\partial_w F_D(y,w) = \int \frac{\mathrm{d}^D
  \mathitbf{u}}{(2\pi)^D}\,\mathrm{e}^{\mathrm{i}\mathitbf{u}\cdot\mathitbf{y}}
\mathrm{e}^{-w(u^2+1)} = \frac{1}{(4\pi)^{D/2}}
\frac{\mathrm{exp}[-w -\frac{y^2}{4w}]}{w^{D/2}}\,.
\end{equation}
Finally, because $F_D(y,w=0)=0$, Eq.~\eqref{eq:FF-representacion-integral} is
recovered from
\begin{equation}
F_D(y,w)=F_D(y,w)-F_D(y,w=0)=\int_0^w\mathrm{d}s\, \partial_s F_D(y,s)\,.
\end{equation}

\section*{References}

\bibliographystyle{iopart-num} 

\end{document}